\documentstyle[aps,prl,twocolumn,epsf,floats]{revtex}
\draft
\begin{document}
\wideabs{

\title{Spin Instabilities and Quantum Phase Transitions\\
       in Integral and Fractional Quantum Hall States}

\author{Arkadiusz W\'ojs$^{1,2}$ and John J. Quinn$^1$}

\address{
   $^1$Department of Physics,
       University of Tennessee, Knoxville, Tennessee 37996, USA\\
   $^2$Institute of Physics,
       Wroclaw University of Technology, Wroclaw 50-370, Poland}

\maketitle

\begin{abstract}
The inter-Landau-level spin excitations of quantum Hall states at
filling factors $\nu=2$ and ${4\over3}$ are investigated by exact
numerical diagonalization for the situation in which the cyclotron
($\hbar\omega_c$) and Zeeman ($E_{\rm Z}$) splittings are comparable.
The relevant quasiparticles and their interactions are studied,
including stable spin wave and skyrmion bound states.
For $\nu=2$, a spin instability at a finite value of $\varepsilon
=\hbar\omega_c-E_{\rm Z}$ leads to an abrupt paramagnetic to
ferromagnetic transition, in agreement with the mean-field
approximation.
However, for $\nu={4\over3}$ a new and unexpected quantum phase
transition is found which involves a gradual change from paramagnetic
to ferromagnetic occupancy of the partially filled Landau level
as $\varepsilon$ is decreased.
\end{abstract}
\pacs{73.43.Nq, 75.30.Fv, 73.43.-f, 73.21.-b}
}

The elementary excitations of a two-dimensional electron gas (2DEG)
with energy quantized into Landau levels (LL's) by a high magnetic
field $B$ have been extensively studied for decades.
The charge excitations govern transport, including the integral and
fractional quantum Hall effects (IQHE and FQHE) \cite{Prange87}.
The spin excitations appear in the context of spin waves (SW's)
\cite{Kallin84}, spin instabilities and related quantum phase
transitions (QPT's) \cite{Giuliani84,Koch93}, and skyrmions
\cite{Lee90,skyrmion}.

\begin{figure}
\epsfxsize=3.40in
\epsffile{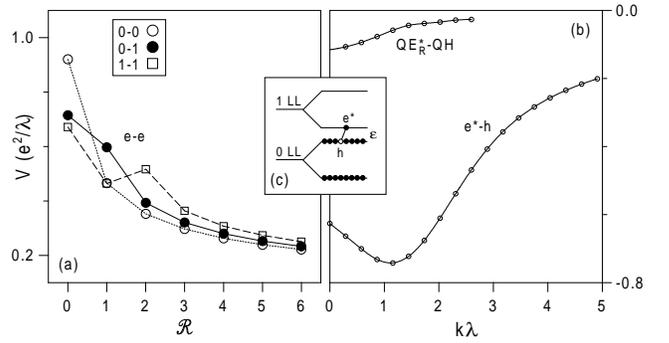}
\caption{
   The Coulomb pseudopotentials $V$ for the pair of:
   (a) electrons in the $n=0$ and 1 LL's, and
   (b) reversed-spin electron ($e^*$) or quasielectron
   (QE$_{\rm R}^*$) in the $n=1$ LL and hole ($h$) or
   quasihole (QH) in the $n=0$ LL.
   (c) Schematic of the LL structure at $\nu=2$, with the $h$
   and $e^*$ quasiparticles.
}
\label{fig1}
\end{figure}

In this letter we study spin excitations of IQH and FQH systems with
densities $\varrho$ corresponding to the filling factors $\nu\!=\!
2\pi\varrho\lambda^2\!\approx\!2$ and ${4\over3}$ (here, $\lambda=
\sqrt{\hbar c/eB}$ is the magnetic length).
The cyclotron ($\hbar\omega_c$) and Zeeman ($E_{\rm Z}$) splittings
are assumed comparable and much larger than the Coulomb energy
$E_{\rm C}=e^2/\lambda$.
In this situation, the spin excitations couple two partially
filled LL's with different orbital indices, $n=0$ and 1.
These LL's, denoted by $\left|0\!\uparrow\right>$ and $\left|1\!
\downarrow\right>$, are separated by a small gap $\varepsilon=\hbar
\omega_c-E_{\rm Z}\ll E_{\rm C}$ from each other and by large gaps
$\sim\hbar\omega_c\gg E_{\rm C}$ from the lower, filled
$\left|0\!\downarrow\right>$ LL and from the higher, empty LL's,
as shown schematically in Fig.~\ref{fig1}(c).

For the $\nu=2$ ground state (GS), it is well-known \cite{Giuliani84}
that a spin-flip instability occurs at a finite gap $\varepsilon$
and wave vector $k$.
In the mean-field approximation (MFA), this instability signals
an abrupt, interaction-induced QPT from paramagnetic (P; $\left|
0\!\downarrow\right>$ and $\left|0\!\uparrow\right>$ filled) to
ferromagnetic (F; $\left|0\!\downarrow\right>$ and $\left|1\!
\downarrow\right>$ filled) occupancy.
Our numerical results confirm the validity of the MFA for $\nu=2$.
However, for $\nu={4\over3}$ they predict a new and unexpected
P$\rightarrow$F QPT that occurs through a series of intermediate
GS's involving increasing number of spin flips as $\varepsilon$
is decreased from $\varepsilon_{\rm P}$ to $\varepsilon_{\rm F}$
(the lower and upper boundaries of $\varepsilon$ for the P and F
occupancies, respectively).

The model is the same as that used earlier \cite{skyrmion,qer},
except that now the spin excitations connect two different LL's.
The electrons are confined to a spherical surface \cite{Haldane83}
of radius $R$.
The radial magnetic field $B$ is due to a monopole of strength
$2Q$, defined in units of the flux quantum $\phi_0=hc/e$ so that
$4\pi R^2B =2Q\phi_0$ and $R^2=Q\lambda^2$.
The single-electron states are labeled by angular momentum $l=Q+n$
and its projection $m$.

Only the partially filled $\left|0\!\uparrow\right>$ and $\left|1\!
\downarrow\right>$ LL's (labeled by pseudospin $s=\,\uparrow$ and
$\downarrow$) are included in the calculation, and the filled,
rigid $\left|0\!\downarrow\right>$ LL enters through the exchange
energy $\Sigma_{10}$.
The ratio $\varepsilon/E_{\rm C}$ is taken as an arbitrary parameter.
Although we do not discuss the effect of the finite width $w$ of a
realistic 2DEG \cite{skyrmion} and only present the results obtained
using the pseudopotential $V({\cal R})$ (interaction energy as a function
of relative pair angular momentum \cite{Haldane87}) for $w=0$, shown
in Fig.~\ref{fig1}(a), we have checked that our conclusions remain
valid for $w\le5\lambda$.

The Hamiltonian $H$ for electrons confined to the $\left|0\!
\uparrow\right>$ and $\left|1\!\downarrow\right>$ LL's contains
the single-particle term $(\varepsilon-\Sigma_{10})$ and the intra-
and inter-LL two-body interaction matrix elements $\left<m_1s,m_2s'
|V|m_3s',m_4s\right>$ calculated for the Coulomb potential $V(r)=
e^2/r$ and connected with pseudopotentials $V_{ss'}({\cal R})$
shown in Fig.~\ref{fig1}(a) through the Clebsch--Gordan coefficients
(on a sphere, ${\cal R}=2l-L$ where $L={\bf l}_1+{\bf l}_2$ is pair
angular momentum).

Hamiltonian $H$ is diagonalized in the basis of $N$-electron
Slater determinants $\left|m_1s_1 \dots m_Ns_N\right>$.
This allows automatic resolution of the projection of pseudospin
($S_z=\sum s_i$) and of angular momentum ($L_z=\sum m_i$).
The quantum number $K={1\over2}N+S_z$ measures the number of
reversed spins relative to the paramagnetic configuration.
The length of angular momentum ($L$) is resolved numerically
in the diagonalization of each $(S_z,L_z)$ Hilbert subspace.
The length of pseudospin is not a good quantum number because
of the pseudospin-asymmeteric interactions.
The results obtained on Haldane sphere are easily converted
to the planar geometry, where $L$ and $L_z$ are appropriately
\cite{geometry} replaced by the total and center-of-mass angular
momentum projections, $M$ and $M_{\rm CM}$.

\begin{figure}
\epsfxsize=3.40in
\epsffile{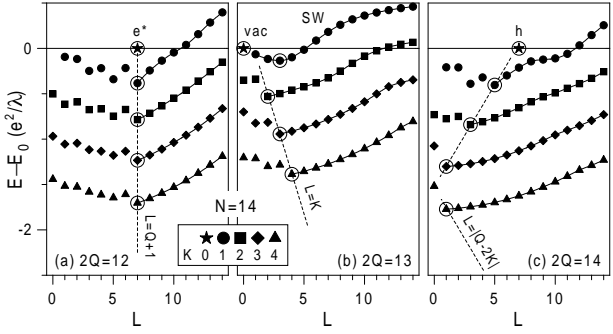}
\caption{
   The excitation energy spectra (energy $E$ as a function
   of angular momentum $L$) of $N=14$ electrons in the
   $\left|0\!\uparrow\right>$ and $\left|1\!\downarrow\right>$
   LL's calculated on a sphere for the monopole strengths $2Q=12$
   (a), 13 (b), and 14 (c), corresponding to the filling factors
   $\nu\approx2$.
   The lowest $\left|0\!\downarrow\right>$  LL is filled.
   $E_0$ is the energy of the lowest paramagnetic ($K=0$) state,
   and dashed lines mark the lowest states for different values
   of $K$.
}
\label{fig2}
\end{figure}

Let us begin with the discussion of the IQH regime.
Fig.~\ref{fig2} presents the spin-excitation spectra for $N=14$,
at the filling factors equal to or different by one flux from
$\nu=2$.
Only the lowest state is shown for each $K$ and $L$.
The energy $E$ is measured from the lowest paramagnetic state
(at $E=E_0$) and excludes the inter-LL gap $\varepsilon$.
Symbols $e^*$ and $h$ denote reversed-spin electrons (particles
in the $\left|1\!\downarrow\right>$ LL) and holes (vacancies in
the $\left|0\!\uparrow\right>$ LL) created in the ``vacuum''
state (completely filled $\left|0\!\uparrow\right>$ LL).

The excitation spectrum of the ``vacuum'' state is shown in
Fig.~\ref{fig2}(b).
The $K=1$ band is a SW; in a finite system it has $L=1$ to $N$,
as follows from addition of the $e^*$ and $h$ angular momenta,
$l_{e^*}=Q+1$ and $l_h=Q$.
In an infinite system, the continuous SW dispersion is given by
\cite{Kallin84} $E_{\rm SW}(k)=E_0+{1\over2}E_{\rm C}\sqrt{\pi/2}
\,\{1-\exp(-\kappa^2)[(1+2\kappa^2)I_0(\kappa^2)-2\kappa^2I_1
(\kappa^2)]\}$, where $\kappa={1\over2}k\lambda$, $I_0$ and $I_1$
are the modified Bessel functions, and $k=L/R$.
$E_{\rm SW}(k)$ starts at $E=E_0$ for $k=0$ and has a minimum at
$k\approx1.19\lambda^{-1}$ and $E\approx E_0-0.147\,E_{\rm C}$.
The vanishing of SW energy at $k=0$ is the result of exact
cancellation of the sum of $e^*$ and $h$ exchange self-energies,
$-\Sigma_{10}+\Sigma_{00}$, by the $e^*$--$h$ attraction $V_{e^*h}$
at $k=0$; the entire $e^*$--$h$ pseudopotential is shown in
Fig.~\ref{fig1}(b).

The energy spectra corresponding to consecutive spin flips
($K=2$, 3, \dots) at $\nu\!=\!2$ all contain low-energy bands at
$L\ge K$.
For each $K$, the GS's (open circles) have $L=K$ and their
energies fall on a nearly straight line, $E(K)$.
These GS's are therefore denoted by ${\cal W}_K=K\times{\rm SW}$
and interpreted as containing $K$ SW's with parallel angular
momenta each of length $L=1$, similar to the $L=K$ SW condensates
at $\nu=1$ \cite{skyrmion}.
The new feature at $\nu=2$ is the SW--SW attraction (due to a
finite dipole moment of an inter-LL SW) giving rise to a negative
slope of $E(K)$.

Let us now turn to Fig.~\ref{fig2}(a) and (b) showing spin
excitation spectra in the presence of an $e^*$ or $h$.
The series of GS's for $K\ge1$ (open circles) are charged bound
states, similar to the skyrmions and anti-skyrmions at $\nu=1$.
Their angular momenta result from simple vector addition of
$l_{e^*}$ and $l_h$.
For ${\cal S}_K^-=K\times{\rm SW}+e^*$ and ${\cal S}_K^+=K\times
{\rm SW}+h$ we get $L=({\bf l}_{e^*})^{K+1}\oplus({\bf l}_h)^K=Q+1$
and $L=({\bf l}_{e^*})^K\oplus({\bf l}_h)^{K+1}=|Q-2K|$, respectively.
In both cases, finite $L\propto Q$ means massive LL degeneracy,
as expected for charged particles in a magnetic field.

\begin{figure}
\epsfxsize=3.40in
\epsffile{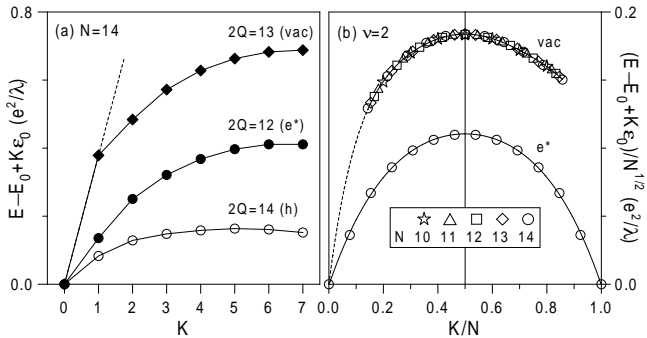}
\caption{
   (a) The energy of skyrmions, anti-skyrmions, and spin-wave
   condensates of Fig.~\protect\ref{fig2}, plotted as a function
   of $K$.
   The gap $\varepsilon$ is set to the value $\varepsilon_0$
   at which the paramagnetic ($K=0$) and ferromagnetic ($K=N$)
   configurations are degenerate.
   (b) The energy of spin-wave condensates calculated for $N=10$
   to 14, rescaled by $\protect\sqrt{N}$, and plotted as a function
   of $\zeta=K/N$.
   The skyrmion curve is shown for comparison.
}
\label{fig3}
\end{figure}

Let us check if the negative SW energy at $k\approx1.19\lambda^{-1}$
or the SW--SW attraction causes instability of the $\nu=2$ GS towards
the formation of one or more SW's when $\varepsilon$ is decreased.
The single-SW instability has been ruled out by Giuliani and Quinn
\cite{Giuliani84} who showed that it is preempted by a direct
transition to the ferromagnetic GS.
The critical value of $\varepsilon$ for this P$\rightarrow$F
QPT is expressed through the involved self-energies,
$\varepsilon_0=\Sigma_{10}+{1\over2}(\Sigma_{11}-\Sigma_{00})=
{3\over8}\sqrt{\pi/2}\,E_{\rm C}\approx0.47\,E_{\rm C}$,
and it is larger than $E_0-E_{\rm SW}$.
Since the energy per spin flip, $[E(K)-E_0]/K$, is smaller for
the SW condensates and skyrmions than for a single SW, we still
need to check for a possible vac$\rightarrow$${\cal W}_K$,
$e^*$$\rightarrow$${\cal S}_K^-$, or $h$$\rightarrow$${\cal S}_K^+$
instability.
Fig.~\ref{fig3}(a) shows that despite evident SW--SW, SW--$e^*$,
and SW--$h$ attraction ($\delta E=E-E_0+K\varepsilon_0$ is the
energy to create $K$ SW's in ``vacuum'' or in the presence
of an $e^*$ or $h$), the ${\cal W}_K$ and ${\cal S}_K^\pm$
energies are all positive at $\varepsilon=\varepsilon_0$.
This precludes spin instability at $\nu=2$ other than the direct
P$\rightarrow$F transition (skipping the states with intermediate
spin).

To translate our finite-size spectra to the case of an infinite
2DEG, in Fig.~\ref{fig3}(b) we have plotted the energies of the SW
condensate calculated for different electron numbers, $N\le14$.
Clearly, all data fall on the same curve when $\delta E/\sqrt{N}$
is plotted as a function of ``relative'' spin polarization,
$\zeta=K/N$.
This resembles the insensitivity to $N$ of the $\delta E(\zeta)$
curves for the SW condensates at $\nu=1$, except that now $\delta
E\propto N^{1/2}$ (rather than $\propto N^0$).

The data of Fig.~\ref{fig3} allows calculation of the SW binding
energies, $U_K=[E(K-1)-E_0]+[E_{\rm SW}-E_0]-[E(K)-E_0]$, for the
${\cal W}_K$ and ${\cal S}_K^\pm$ states.
Because of the SW--SW attraction, all these energies increase in
a similar way as a function of $K$, in contrast to $\nu=1$ where
$U_K$ decreased for skyrmions and vanished for the SW condensate.

Let us now turn to the FQH regime.
At $\nu={4\over3}$, which occurs for $2Q=3(N-1)$, and for
sufficiently large $\varepsilon$, the $N$ electrons in the
$\left|0\!\uparrow\right>$ LL form the Laughlin $\nu={1\over3}$
state.
These electrons, each with angular momentum $l=Q$, can be
converted into an equal number of composite fermions (CF's)
\cite{Jain89} each with effective angular momentum $l^*=l-(N-1)$,
exactly filling their effective LL.
The elementary charge excitations of the $\nu={1\over3}$ state
are two types of Laughlin quasiparticles (QP's), quasielectrons
(QE's) and quasiholes (QH's), corresponding to an excess particle
in an (empty) excited CF LL, or a hole in the (filled) lowest CF
LL, respectively.

The reversed-spin quasielectrons (QE$_{\rm R}$'s) \cite{qer,Rezayi87}
do not occur at $\nu={4\over3}$ because of the electrons completely
filling the $\left|0\!\downarrow\right>$ LL.
This causes a difference between the SW's at $\nu={4\over3}$ and
${1\over3}$, similar to that between $\nu=2$ and 1.
At $\nu={1\over3}$ the SW consisted of a QH and a QE$_{\rm R}$,
and at $\nu={4\over3}$ it is formed by a QH and a different
reversed-spin QP that we will denote by QE$_{\rm R}^*$.

The QE$_{\rm R}^*$ has the same electric charge of $-{1\over3}e$
as QE or QE$_{\rm R}$ but it belongs to an excited electron LL,
$\left|1\!\downarrow\right>$.
Similar to the case for QH, QE, and QE$_{\rm R}$, the existence
and stability of the QE$_{\rm R}^*$ depend on the validity of the
CF transformation for the underlying system of $N-1$ electrons in
the $\left|0\!\uparrow\right>$ LL and one electron in the $\left|
1\!\downarrow\right>$ LL.
This requires Laughlin correlations between the $\left|1\!\downarrow
\right>$ electron and the $\left|0\!\uparrow\right>$ electrons,
i.e.\ the occurrence of a Jastrow prefactor, $\prod_{ij}(z^{(0)}_i-
z^{(1)}_j)^\mu$, in the many body wave function, with $\mu=2$ for
$\nu=(1+\mu)^{-1}={1\over3}$.
Such correlations result from short-range $e$--$e$ repulsion,
and the criterion is \cite{jphys,fivehalf} that the pseudopotential
$V$ must decrease more quickly than linearly as a function of the
average square $e$--$e$ separation $\left<r^2\right>$.
On a plane (or on a sphere for $\left<r^2\right>\ll R^2$, i.e.\ for
${\cal R}\ll Q$) this is equivalent to a superlinear
decrease of $V$ as a function of ${\cal R}$.

It is clear from Fig.~\ref{fig1}(a) that the Coulomb inter-LL
pseudopotential $V_{01}({\cal R})$ is a short-range repulsion
for ${\cal R}\ge{\cal R}_0=1$.
This implies the Jastrow prefactors with $\mu>{\cal R}_0=2$, 3,
\dots\ in the $\left|0\!\uparrow\right>^{N-1}\oplus\left|1\!
\downarrow\right>$ wave function, if only $\nu\le(1+\mu)^{-1}$.
In particular, this establishes the QE$_{\rm R}^*$ as a stable
reversed-spin QP of the $\nu={4\over3}$ state, in analogy to
the reversed-spin electron, $e^*$, at $\nu=2$.
The angular momentum of QE$_{\rm R}^*$ on a sphere can be obtained
in the two-component CF picture \cite{x-cf} appropriate for $\nu=
{1\over3}$, i.e.\ with both 0--0 and 0--1 Laughlin correlations
modeled by attachment of two flux quanta to each electron.
The resulting CF angular momenta are $l_{\rm QH}=Q^*$ and
$l_{\rm QE}=l_{{\rm QE}_{\rm R}^*}=Q^*+1$, where $Q^*=Q-(N-1)$.

\begin{figure}
\epsfxsize=3.40in
\epsffile{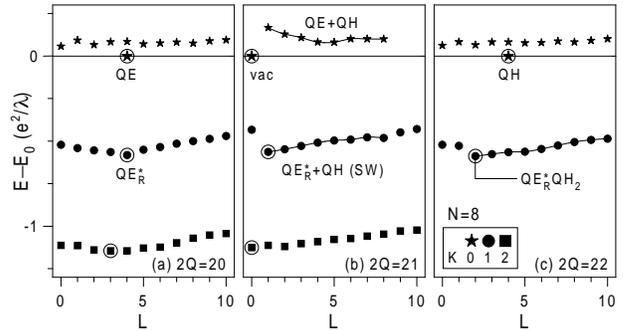}
\caption{
   Same as Fig.~\protect\ref{fig2}, but for $N=8$ electrons and
   for the monopole strengths $2Q=20$ (a), 21 (b), and 22 (c),
   corresponding to the filling factors $\nu\approx{4\over3}$.
}
\label{fig4}
\end{figure}

The excitation spectra at filling factors equal to or different
by one flux from $\nu={4\over3}$ are displayed in Fig.~\ref{fig4}.
$N=8$ in each frame, and the values of $2Q$ are 20, 21, and 22,
corresponding to the following GS's at $K=0$: (a) QE at $L=4$,
(b) ``vacuum'' (filled CF LL) with $L=0$, and (c) QH at $L=4$.
The low-energy charge excitations for $2Q=21$ form the magnetoroton
(QE$+$QH) band.
The low-energy spin excitations with $K=1$ are the following:
(a) QE$_{\rm R}^*$ at $L=l_{{\rm QE}_{\rm R}^*}=4$ for $2Q=20$,
(b) the SW (QE$_{\rm R}^*$$+$QH) band with $L$ going from 1 to
$N=8$, as follows from vector addition of $l_{\rm QH}$ and
$l_{{\rm QE}_{\rm R}^*}$, for $2Q=21$, and
(c) a band of QE$_{\rm R}^*$$+$QH$_2$ states with a bound
GS denoted as QE$_{\rm R}^*$QH$_2$ for $2Q=22$.

To draw analogy with Fig.~\ref{fig2}, QE corresponds to an
electron in the $\left|1\!\uparrow\right>$ LL (not shown because
of high energy), QE$_{\rm R}^*$ to $e^*$, QH to $h$, and
QE$_{\rm R}^*$QH$_2$ to ${\cal S}_1^+$.
The latter state is the only ``skyrmion'' at $\nu={4\over3}$.
The ${\cal S}_K^-$ states with $K\ge1$ and $L=Q^*+1$ or the
${\cal S}_K^+$ states with $K\ge2$ and $L=|Q^*-2K|$ do not
occur because of the weakened Coulomb repulsion at short range
in the excited LL.
As shown in Fig.~\ref{fig1}(a), the linear behavior of
$V_{11}({\cal R})$ between ${\cal R}=1$ and 5 prevents
Laughlin correlations for two or more electrons in the $n=1$ LL.
This invalidates the CF model and causes break-up of
QE$_{\rm R}^*$'s when two of them approach each other
(at this point, pairing of electrons in the $n=1$ LL occurs
\cite{fivehalf,Moore91}).
For the same reason, no ${\cal W}_K$ states at $L=K$ appear
in Fig.~\ref{fig4}(b) for $K>1$.

Even more significant in Fig.~\ref{fig4} than the absence of
${\cal S}_K^\pm$ and ${\cal W}_K$ states is the large and
negative SW energy $E_{\rm SW}^*(k)$ at $\nu={4\over3}$.
This is in striking contrast to the $\nu=2$ case,
and it is explained as follows.
The SW energy is the sum of the QE$_{\rm R}^*$ and QH
self-energies and the QE$_{\rm R}^*$--QH attraction.
Of these three terms, only the QE$_{\rm R}^*$ self-energy,
$-\Sigma_{10}=-{1\over2}\sqrt{\pi/2}\,E_{\rm C}$, is the
same at $\nu=2$ and ${4\over3}$, while the QH self-energy
$\Sigma_{00}^*$ and the QE$_{\rm R}^*$--QH pseudopotential
$V_{{\rm QE}_{\rm R}^*{\rm QH}}(k)$ are both reduced (because
of only partial filling of the $\left|0\!\uparrow\right>$ LL
and the fractional QP charge, respectively).
As a result, the large and negative $-\Sigma_{10}$ term becomes
dominant in $E_{\rm SW}^*(k)$.
Note that even without knowing analytic expressions for
$\Sigma_{00}^*$ or $V_{{\rm QE}_{\rm R}^*{\rm QH}}(k)$, the
fact that $V_{{\rm QE}_{\rm R}^*{\rm QH}}(\infty)=0$ allows
the estimate of $V_{{\rm QE}_{\rm R}^*{\rm QH}}(k)$, as shown in
Fig.~\ref{fig1}(b), and of $\Sigma_{00}^*\approx0.17\,E_{\rm C}$.
Note that $V_{{\rm QE}_{\rm R}^*{\rm QH}}(0)\approx-0.11\,
E_{\rm C}\approx{1\over6}V_{e^*h}(0)$ and $\Sigma_{00}^*\approx
{1\over7}\Sigma_{00}$.

\begin{figure}
\epsfxsize=3.40in
\epsffile{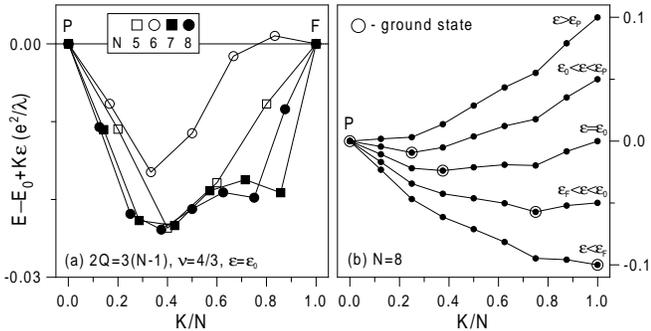}
\caption{
   (a) Same as Fig.~\protect\ref{fig3}(b), but for the filling
   factor $\nu={4\over3}$.
   (b) Data for $N=8$ plotted for different values of $\varepsilon$.
}
\label{fig5}
\end{figure}

The dependence of the GS energy on $\zeta=K/N$ for $\nu={4\over3}$
is shown in Fig.~\ref{fig5}(a).
As in Fig.~\ref{fig3}, $\varepsilon$ is set to the value
$\varepsilon_0$ for which the P and F configurations
(at $\zeta=0$ and 1) are degenerate.
Clearly, (almost) all energies at $0<\zeta<1$ are negative.
This effect does not depend on $N$; on the contrary, all data points
for moderate values of $\zeta$ seem to to fall on the same curve,
characteristic of an infinite (planar) system.
Negative excitation energies imply that the paramagnetic Laughlin
$\nu={4\over3}$ state is unstable toward flipping of only a fraction
$\zeta<1$ of spins when $\varepsilon$ is decreased.
This is illustrated in Fig.~\ref{fig5}(b) where we display the data
for $N=8$ corresponding to five different values of $\varepsilon$.
The gradual decrease of $\varepsilon$ from $\varepsilon_{\rm P}$ to
$\varepsilon_{\rm F}$ drives the system through entire series of GS's
(open circles) with fractional values of $\zeta$.
This novel sequence of GS's are distinctly different from the abrupt
P$\rightarrow$F QPT found at $\nu=2$, and they are not expected in
the MFA.

In conclusion, our numerical study of small systems at $\nu=2$ serves
as a test of the MFA which predicts an abrupt interaction-induced
P$\rightarrow$F QPT associated with the spin-flip instability.
This test should also be applicable to a similar instability and
QPT which occurs for a bilayer \cite{DasSarma94} (where
$\hbar\omega_c$ is replaced by the symmetric-antisymmetric splitting
$\Delta_{\rm SAS}$).
For the fractional $\nu={4\over3}$ state the series of spin-flip
GS's between the para- and ferromagnetic states is a novel
prediction that is susceptible to experimental observation.

The authors acknowledge partial support by the Materials Research
Program of Basic Energy Sciences, US Department of Energy, and
thank I. Szlufarska and G. Giuliani for helpful discussions.

\end{document}